\documentclass[11pt]{article}
\usepackage{times}
\usepackage{geometry}
\usepackage[utf8]{inputenc}
\usepackage{enumitem,amssymb}
\usepackage{graphicx}
\usepackage{fancyhdr}
\usepackage{aas_macros}
\usepackage{mdframed} 
\usepackage{xcolor}

\usepackage[authoryear]{natbib}
\setcitestyle{authoryear,open={(},close={)}}

\mdfdefinestyle{theoremstyle}{
innertopmargin=\topskip,}
\mdtheorem[style=theoremstyle]{lrptextbox}{}

\begin{document}
\title{The Formation of Stars - From Filaments to Cores \\
to Protostars and Protoplanetry Disks}

\author{James Di Francesco (NRC), Helen Kirk (NRC), Doug Johnstone (NRC), \\
Ralph Pudritz (McMaster), Shantanu Basu (Western), Sarah Sadavoy (Queen's), \\
Laura Fissel (Queen's), Lewis Knee (NRC), Mehrnoosh Tahani (NRC), \\
Rachel Friesen (Toronto), Simon Coud\'e (SOFIA-USRA), Erik Rosolowsky (Alberta), \\
Nienke van der Marel (NRC), Michel Fich (Waterloo), Christine Wilson (McMaster), \\
Chris Matzner (Toronto), Ruobing Dong (Victoria), Brenda Matthews (NRC), \\
and Gerald Schieven (NRC)}
\maketitle

\begin{abstract}
Star formation involves the flow of gas and dust within molecular clouds into protostars and young stellar objects (YSOs) due to gravity. Along the way, these flows are shaped significantly by many other mechanisms, including pressure, turbulent motions, magnetic fields, stellar feedback, jets, and angular momentum. How all these mechanisms interact nonlinearly with each other on various length scales leads to the formation and evolution of substructures within clouds, including filaments, clumps, cores, disks, outflows, the protostars/YSOs themselves, and planets. In this white paper, prepared for the 2020 Long Range Plan panel which will recommend Canada's future directions for astronomy, we describe the observational and theoretical leadership in the star formation field that Canada's vibrant community has demonstrated over the past decade. Drawing from this extensive background, we identify five key questions that must be addressed for further progress to be made in understanding star formation in the next decade. Addressing these questions will improve our understanding of the dynamics of the dense gas and the role of the magnetic field in star formation, the optical properties of the dust used to trace mass and magnetic fields, the sources of variability in star-forming objects on short timescales, and the physical processes that specifically promote the clustering of stars. We further highlight key facilities in which Canada should become involved to continue making progress in this field. Single-dish facilities we recommend include LSST, trans-atmospheric far-infrared telescopes like BLAST-TNG and SPICA, and ground-based telescopes like JCMT, GBT, and CCAT-p. Interferometric facilities we recommend include ALMA, ngVLA, and SKA1.
\end{abstract}


%


~

\section{Star Formation Research in Canada}

Stars are among the most fundamental constituents of the baryonic Universe.  They populate galaxies, produce heavy elements, and warm the surfaces of exoplanets.  Given the importance of stars, it is also important to understand their origins.  Indeed, our own origins 
are directly connected to the formation of our Sun five billion years ago.  Accordingly, star formation is an active part of modern astrophysics research and Canadians have been involved in it for decades. In this LRP2020 white paper, we describe the recent activities of Canada’s vibrant star formation community, summarize the important questions that need to be addressed in the coming decade, and make recommendations for current and future facilities best suited to answering those questions.  

Star formation is a fascinating process because it involves a very wide range of length scales and physical mechanisms.  For example, a typical molecular cloud is 10s of~pc in size, but the denser, star-forming substructures like clumps or cores within clouds are only 1~pc or 0.1~pc in size, respectively.  Furthermore, protoplanetary disks around young stellar objects (YSOs) that form within cores are about 0.001~pc in size, and the YSOs themselves are about 10$^{-8}$~pc (or less) in size.  Most stars form as members of clusters within overdense regions in molecular clouds known as clumps.  The most massive cluster forming clumps form at the intersections (or "hubs") of systems of filaments.  Gas flows along these filaments bring fresh material into the star formation cauldron that is a young cluster.  To form an individual star, gravity ultimately draws diffuse material (gas and dust) from a molecular cloud into a denser core and then onto a disk and into a YSO.  The inward pull of gravity, however, is countered by the outward push of thermal pressure, and this push/pull is further affected by stellar feedback, protostellar outflows and jets, turbulence, magnetic fields, and angular momentum.  The final mass that a star attains is likely a consequence of how protostellar outflows and radiation fields act on the disk and collapsing gas to halt the inflow.  Indeed, the challenge of star formation is to constrain the interplay between these various mechanisms, which can act nonlinearly and have different degrees of importance across the length scales involved.  Ultimately, we would like to have an accurate predictive theory of star formation, where we could say with confidence how many stars and planets (and what types of each) are produced by a given parcel of gas in a galaxy.

On the theoretical side, star formation is difficult to express analytically due to the complex behaviors of the physical mechanisms involved.  Instead, simulations generally incorporate subsets of these mechanisms and include assumptions on the initial and boundary conditions, dimensions, and scales involved to make their executions tractable. 
Fortunately, the rise of ever more powerful computers in the last few decades has made more comprehensive simulations possible, and which now include the effects of radiative feedback from forming massive clusters and stars, as well as magnetic fields.  On the observational side, star formation is difficult to probe visually due to the large amounts of extincting dust within molecular clouds.  Instead, data about forming stars require observations of wavelengths longer than the visible range that are less susceptible to extinction, though they come at a cost of a lower angular resolution for a given telescope size (i.e., $\sim$ $\lambda$/D).  Fortunately, access to wavelengths from the near-infrared (0.001 mm) to the far-infrared (0.1 mm) and millimetre (3 mm) has opened up considerably in the last few decades due to new technologies (i.e., telescopes, detectors, and receivers).  Over these wavelengths, thermal emission from dust and gas is actually bright both in continuum and lines, respectively, from warm circumstellar material (100-1000 K) to cold (10-100 K) molecular clouds.  Furthermore, significant improvements in resolution have been achieved due to other new technologies (i.e., larger apertures, interferometry). 

Theorists and observers have worked together to make the most impactful progress in star formation.  In the past decade, Canadians have harnessed new computational and observational resources to make significant inroads. 
This white paper will focus on star formation within our own Galaxy, and mostly on aspects between the disk and molecular cloud scales, where 
the cores and clumps that serve as interfaces between clouds and disks/YSOs are studied.  In parallel, 
star and planet formation on scales larger than the cloud and smaller than the disk are described in WP017 (Rosolowsky et al.~2019) and WP005 (van der Marel et al.~2019), respectively.
This separation into different white papers was done mostly to provide adequate descriptions of the issues on the respective scales involved.
A coherent picture of star and planet formation requires information on as wide a range of scales as possible.

\section{The Last Decade}

There has been significant progress made in understanding star formation over the past ten years, mostly thanks to wide-field surveys that have probed star-forming populations on large scales and high-resolution observations that have zeroed in on examples of star formation on small scales.  In this section, we describe these advances in turn, with particular emphasis on results obtained by Canadians.

\subsection{Large-Scale Views}

In the first decade of the 21st century, large programs using the near- to mid-infrared {\it Spitzer} space telescope yielded very complete censuses of the YSO populations of numerous nearby and distant star-forming clouds in our Galaxy  \citep[e.g.,][]{Evans09,Benjamin05,Megeath12,Dunham15}.  These data, tracing YSOs via the warm dust surrounding them in disks and envelopes, provided for the first time very important measures of the timescales involved in YSO evolution
and the variation in YSO populations between clouds. Canadian involvement included work by \cite{Kirk09} and \cite{B-F14}.
On larger scales, observations from  {\it Spitzer} were key to identifying massive star-forming complexes in the outer Galaxy and instances of significant feedback from young high-mass stars generating large expanding bubbles that will eventually disrupt their natal clouds \citep[e.g.,][]{Rahman10,Murray11,Lee12}.

\begin{figure}[ht]
 \centerline{\includegraphics[width=0.7\textwidth]{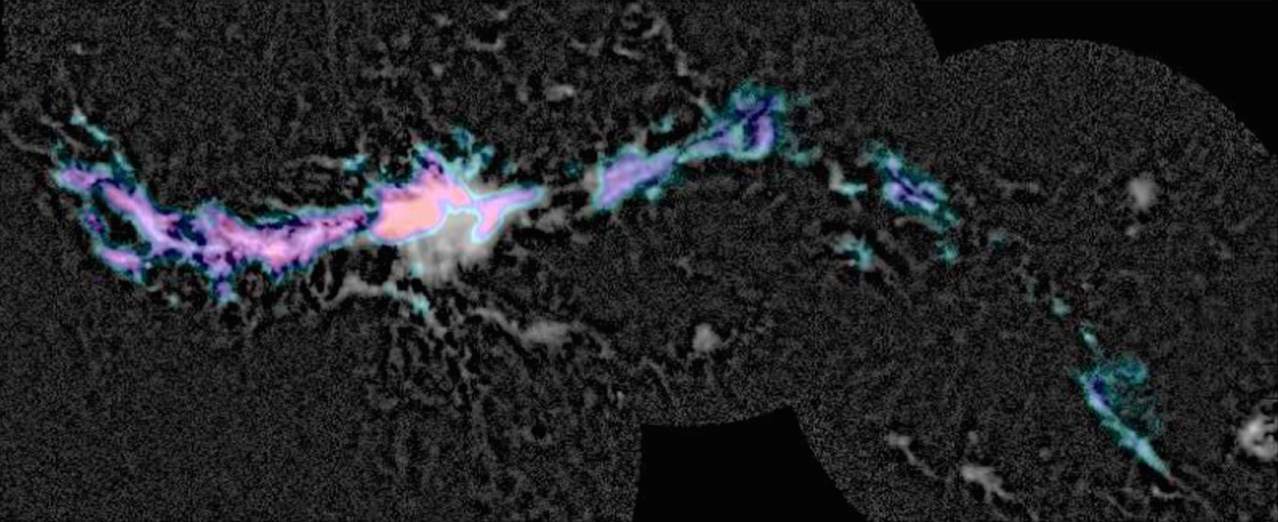}}
 \caption{\label{fig:fig1} 
 850 $\mu$m intensities (greyscale) from the JCMT Gould Belt Survey and NH$_{3}$-derived kinetic temperatures (colorscale) from the Green Bank Ammonia Survey toward Orion A \citep{Kirk17b,Friesen17}. 
 }
\end{figure}

In the last ten years, further wide-field investigations of star-forming clouds continued. Specifically, large programs using the far-infrared/submillimetre {\it Herschel} space observatory mapped thermal continuum emission from cold dust across numerous star-forming clouds both near and far in the Galaxy, revealing the internal structure of these clouds in ways not previously possible from the ground \citep[e.g., see][]{Sadavoy12,Sadavoy14}.  Notably, these observations revealed that molecular clouds are suffused with low column density filamentary structures.  Indeed, earlier simulations had predicted that supersonic turbulence creates a rich tapestry of filaments.  These structures appear to be efficient pathways for the formation of clumps and cores via gravitational fragmentation \citep{Andre10,Fischera12b,Fischera12a}, confirming analytic predictions into their inherent gravitational instabilities \citep{Inutsuka97, Fiege00, Pon11,Pon12b}. Moreover, the {\it Herschel} observations cemented the morphological similarity between core and stellar initial mass functions \citep{Andre14,Konyves15}, although a direct physical connection remains uncertain \citep{Mairs14}.  In addition, large JCMT programs using SCUBA-2, such as the JCMT Gould Belt Survey, mapped several of the same clouds and identified their cold core populations at higher resolutions than possible with {\it Herschel} \citep[e.g.,][]{Kirk16a,Kirk16b,Coude16,Mairs16,Lane16,Johnstone17,B-F18}.  
Figure \ref{fig:fig1} shows an example SCUBA-2 map of the Orion A cloud.  Such data revealed for the first time evidence of mass segregation in clusters of dense cores \citep{Kirk16b,Lane16}, the influence of newly formed massive stars in heating their local surroundings \citep[e.g.,][]{Rumble15,Rumble16}, and 
the increase in YSO velocity dispersions with evolution \citep{Mairs16}.  Furthermore, combinations of the SCUBA-2 and {\it Herschel} data probed dust opacity variations within star-forming regions of clouds, revealing populations of larger dust grains within the intercore medium \citep{Sadavoy13,Chen16}.
Lower-resolution far-infrared/millimetre data acquired by the \emph{Planck} observatory also well constrained the optical properties of dust across the Galaxy, including star-forming clouds.  Those data, however, showed that dust properties are largely similar across many clouds, suggesting that grain evolution proceeds primarily on scales smaller than \emph{Planck} could probe \citep[e.g.,][]{Planck11,Planck14}

With wide-field submillimetre continuum imagers like SCUBA-2, it became possible to monitor star-forming regions over time. During formation, low-mass protostar emission is dominated by accretion luminosity as stellar mass grows \citep{Johnstone13}.  While mass inflow from the envelope may be steady or slowly varying, accretion through the disk and onto the protostar is expected to be highly variable \citep[e.g.,][]{Vorobyov15}. For the past four years, the JCMT Transient Survey \citep{Herczeg17} has monitored eight nearby star-forming regions with a monthly cadence.  It has revealed so far that $\sim$10\% of deeply embedded protostars indeed vary on timescales of months to years \citep{Mairs17,Johnstone18} and uncovered a protostar with an eighteen-month period, possibly the result of influence by a forming planet,  \citep{Yoo17}.  

\begin{figure}[ht]
 \centerline{\includegraphics[width=0.7\textwidth]{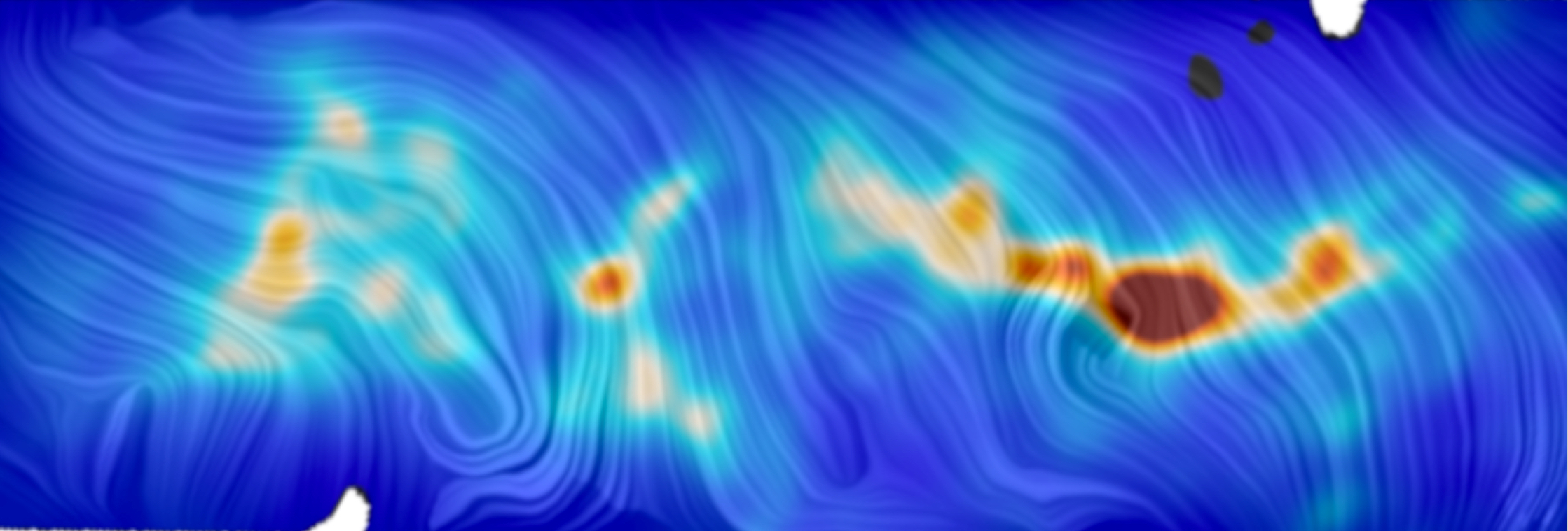}}
 \caption{\label{fig:fig2} 
 Plane-of-sky magnetic field directions associated with the Vela C cloud overlaid onto 500 $\mu$m continuum emission, observed with BLAST-pol by \cite{Fissel16}.  
 }
\end{figure}

Magnetic fields within clouds have been traced extensively over the past decade using maps of polarized far-infrared/submillimetre continuum emission over cloud extents.  Here, polarization of dust emission can be induced, e.g., by the preferential alignment of elongated rotating dust grains within magnetic fields. \emph{Planck} also mapped dust polarization across the entire Galaxy at coarse resolution, revealing the large-scale Galactic magnetic field for the first time. These observations also showed a bimodality between the field orientation and cloud elongation for different cloud densities \citep{Planck15}. Figure \ref{fig:fig2} shows magnetic field directions observed on cloud scales toward the Vela C cloud with the BLAST-pol balloon-borne telescope by \cite{Fissel16,Fissel19}.  Here, similar bimodal orientations were found toward diffuse and dense regions of the cloud, as traced by dust or gas.  In addition, the Canadian-made POL-2 extension to SCUBA-2 provided polarization mapping of individual dense cores, enabling magnetic field morphologies and strengths to be measured and their dynamical importance determined \citep[e.g.,][]{Pattle17,Pattle18,Coude19}.
Indeed, intense work in studying the magnetic fields within star-forming clouds continues at JCMT, but also  elsewhere.  For example, \citet{Tahani18} recently detailed a new method of using Faraday rotation measure and extinction maps to estimate the line-of-sight strengths of magnetic fields associated with molecular clouds. Since this method determines the line-of-sight component of the magnetic field, it can provide hints to the 3D morphology of these fields when combined with other techniques.
Furthermore, \citet{Auddy19} described another new method of determining magnetic field strengths, using instead the turbulent velocity dispersions obtained from wide-field line observations. 

Canadians also obtained wide-field observations of line emission across clouds over the past decade, sampling the molecular gas content of these clouds.  Early large maps of line emission were made using HARP at the JCMT, which sampled more the ambient gas within these clouds \citep[e.g.,][]{Buckle12,Sadavoy13}.  Later studies probed the dense gas more directly involved in star formation \citep[e.g.,][]{Friesen09,Friesen10a}, most recently from wide-field NH$_{3}$ maps made using the KFPA instrument on the GBT \citep{Friesen17,Keown19}. For example, Figure \ref{fig:fig1} also shows kinetic temperatures determined for Orion A using NH$_{3}$ data from the GBT \citep{Kirk17b}.  Such data have elucidated the dynamical states of cores and clumps across clouds, and importantly revealed the previously unappreciated role of external pressure in keeping cores bound when gravity is insufficient \citep{Kirk17b,Keown17,Kerr19}.  In addition, other line data have revealed the motions of gas in filaments, allowing estimates of accretion rates onto clumps \citep[e.g.,][]{Kirk13,Friesen13}.  


Theoretical work on many of these topics also occurred in parallel.  For example, using a numerical simulation of a cluster-forming region, the strong influence of magnetic fields and radiative feedback on the evolution of filaments and their role in supplying material to cluster-forming regions was demonstrated \citep{Kirk15,Klassen17}.  
Also, simulations of feedback from massive cluster-forming regions in Giant Molecular Clouds showed that their masses scale with the host cloud mass, suggesting a universal mechanism of cluster formation, perhaps into the globular cluster regime \citep[e.g.,][]{Howard16,Howard17a,Howard17b,Howard18}.
In addition, studies into star formation efficiencies and rates, as modulated by turbulence, gravity from larger-scale structures, and feedback, were investigated using numerical simulations \citep[e.g.,][]{Murray17,Grudic19}.


\subsection{Small-scale Views}

The last decade has also seen profound advances in the understanding of star formation on scales smaller than that of the core and clump.  These advances are primarily due to access to ALMA and the Jansky VLA, whose unprecedented sensitivities and resolutions from submillimetre to centimetre wavelengths opened up new ways of probing star (and planet) formation.  Here too, Canadians have made significant contributions.

With ALMA, large samples of prestellar cores within molecular clouds could be now observed at high resolution, e.g., to explore the occurrence of smaller-scale structure due to fragmentation or collapse.  By so doing, \cite{Kirk17a} found very little detectable small-scale structure within the starless core population of the nearby Ophiuchus cloud, severely constraining the timescales for the appearance of compact structures in such cores \citep[see also,][]{Dunham16}.  Moreover, star-forming clumps could be now examined at high resolution, to resolve their constituent components.  For example, \cite{Friesen14,Friesen18} used ALMA to probe the youngest objects in the Oph A clump and found the clearest cases so far detected of a ``first hydrostatic core,” a very short-lived ($\sim$1000 yr) phase of protostellar evolution expected prior to the formation of the actual protostar.  Also, ALMA and the Jansky VLA were used to explore comprehensively the origins of multiplicity in highly embedded YSOs for the first time.  Here, \cite{Tobin16,Tobin18} showed statistics of embedded YSOs in the Perseus cloud indicating that close multiple systems, i.e., with separations $<$200 au, clearly arise due to disk fragmentation.  Using these data as well as data of the parent cores from SCUBA-2, \cite{Sadavoy17}  argued that virtually all stars form as multiples, and those that are presently singular became so only through later dynamical interactions. 

Perhaps one of the most important roles that magnetic fields play in star formation is in the launch and collimation of the outflows associated with YSOs of all masses.  Canadians have played a leading role in developing simulations that show that magnetized disk winds remove significant amounts of angular momentum, driving accretion flows onto the protostar \citep[see][]{Pudritz19}.  Recent high-resolution ALMA observations show that disk winds originate from large portions of disks, and that the resulting outflows also rotate \citep{Hirota17,Lee17,Kwon19}.  While magnetically driven turbulence in disks has long been assumed to drive accretion flow onto the star, recent state-of-the-art MHD simulations show that such turbulence is strongly damped, leaving it to the magnetized disk wind to drive stellar accretion flow \citep{Bai13, Gressel15}.  

\begin{figure}[ht]
 \centerline{\includegraphics[width=0.7\textwidth]{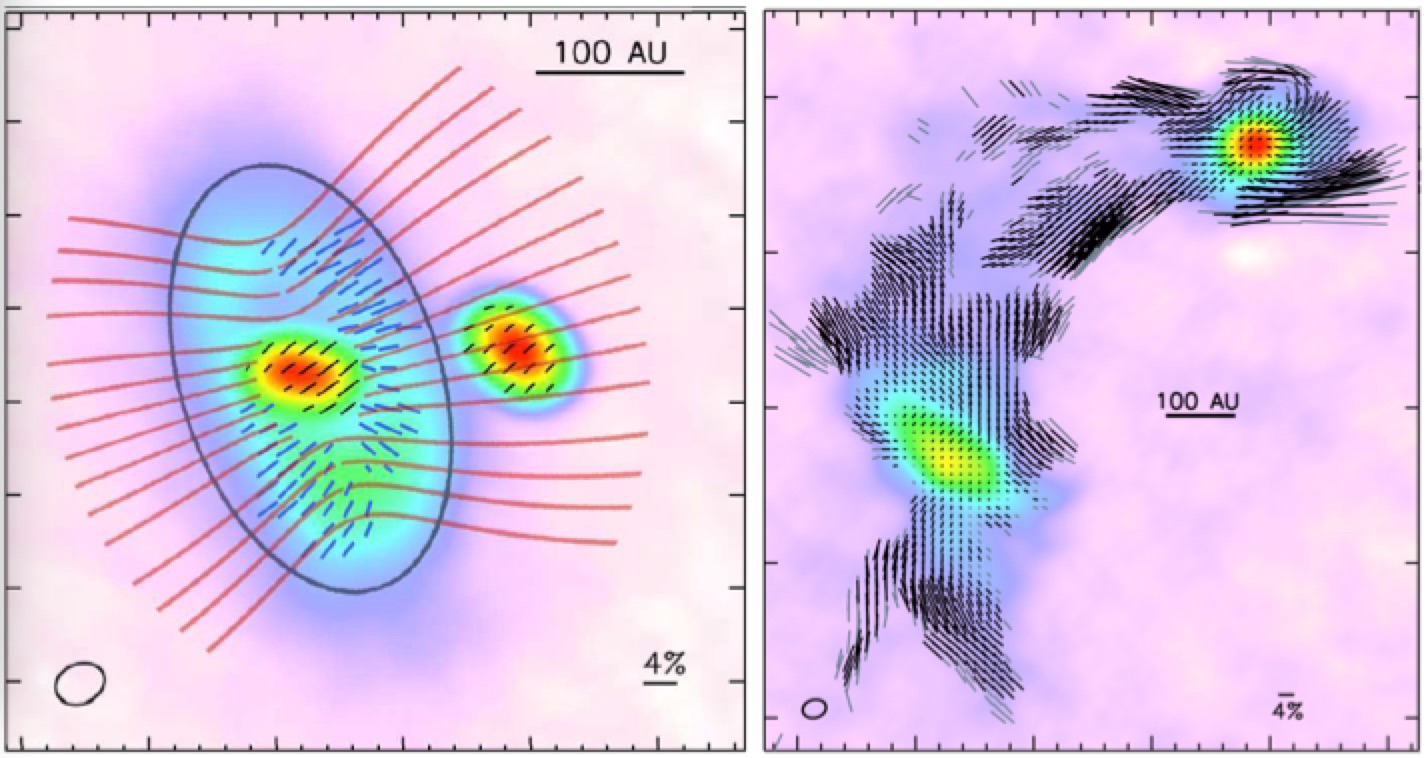}}
 \caption{\label{fig:fig3} 
 Plane-of-sky magnetic field directions overlaid onto 1.3 mm continuum emission from Ophiuchus protostars, each observed with ALMA.  {\it Left panel}: The vectors around VLA1623 (A and B) are consistent with a flux-frozen poloidal field.  The vectors coincident with each protostar, however, are likely due to self-scattering by dust in their disks \citep{Sadavoy18a}. {\it Right panel}: The vectors associated with IRAS 16293-2422 (A and B) show alignment with a bridge of material between the protostars \citep{Sadavoy18b}.}
\end{figure}

The high sensitivities and resolutions of ALMA also enabled for the first time deep probes of polarized emission from regions immediately surrounding YSOs.  Figure \ref{fig:fig3} shows examples from \cite{Sadavoy18b,Sadavoy18a} who probed the structure of polarized emission around embedded YSOs in Ophiuchus.  For VLA 1623 (left panel), the polarization vectors are consistent with a flux-frozen poloidal magnetic field drawn into immediate surroundings of the YSOs during the collapse of the parent core.  Furthermore, these data showed that the polarization signatures of the disk are not consistent with magnetic fields but rather dust scattering, illustrating a major limitation of using such data to trace magnetic fields in the high density material in disks.  For IRAS 16293-2422 (right panel), the polarization data revealed a local magnetic field that runs parallel to a filamentary bridge of material between two YSOs and potentially regulates gas flow between the binary members or accretion channels from the protostellar envelope \citep[e.g.,][]{Seifried15}.


\section{The Big Questions over the next ten years}

With the extensive high-impact research in star formation done by Canadians over the past decade, we are in a good position to identify the big questions that need to be addressed over the next ten years to continue to make progress.

\subsection{How does dense gas flow to produce clumps, cores, and stars?}
 
Though the last ten years have shown that filaments appear critical to produce smaller-scale star-forming structures like clumps and cores, we still know very little about the motions of gas involved in this process.  Continued studies of the dynamics of dense gas are needed to make progress, particularly at higher resolutions.  Single-dish surveys led by Canadians have produced spectacular datasets over the past decade but better detail is needed to associate the dense gas observed with cores identified at higher resolutions via continuum emission.  Specifically, these higher-resolution line data are needed to probe the accretion flows of gas onto filaments, the conditions for fragmentation of filaments that may lead to core formation, and the accretion flows of dense gas along filaments onto cores.  Furthermore, higher spectral resolution data are needed to disentangle multiple structures at different velocities along lines of sight, including within filaments, where recent work suggests that instead of being simple systems which enable accretion, filaments may be composed of multiple separate velocity-coherent structures which are intertwined together \citep[e.g.,][]{Hacar13}. The need for higher resolution is especially acute for probing star-forming structures within Giant Molecular Clouds, which dominate the production of new stars within the Galaxy, as these are much more distant than the less-active clouds closer to us.   
Wide-field observations of the ambient material within molecular clouds are also needed, not necessarily at high resolution, to understand better the role of external pressure in the formation and evolution of filaments, their surrounding magnetic field morphologies, clumps, and cores.  Within cores themselves, high-resolution line data are needed to measure further the accretion flows of gas onto embedded young disks.  

\subsection{What is the role of the magnetic field in the formation and evolution of dense star-forming material?}

Significant advances in probing magnetic fields associated with star formation have been made in the past decade, including 
work by Canadians.  {\it Planck} has revealed a large-scale magnetic field pattern in the Galaxy that implies that magnetic energy is either comparable to or exceeds the turbulent energy on large scales \citep{Planck15}. Meanwhile, 
the distribution of magnetic field morphologies within  clouds is heterogeneous and more difficult to interpret.
Wider samples of field behaviour, over a range of scales and wavelengths, are needed
to probe variations in temperature and dust grain alignment.  On larger scales, such data are also needed to understand the role of the magnetic field in the overall support of molecular clouds and in the formation of filaments.  On smaller scales, wider surveys are also needed to see how magnetic fields behave as cores grow and collapse, and how they promote (or inhibit) further fragmentation, the formation of disks, and the launching of outflows from YSOs.  These data can be combined with line data to estimate field strengths using the Davis-Chandrasekhar-Fermi method \citep[e.g.,][]{Pattle19,Coude19}. These estimates should be also compared to those obtained by other means, e.g., using the Zeeman effect on transitions of the OH or CCS molecules.  Alternatively, newer methods of determining  field strengths that have been pioneered by Canadians can be used \citep[e.g.,][]{Houde11,Houde16,Tahani18,Auddy19}. 

On the theory side, self-consistent modeling of core collapse including magnetic fields and dissipation that leads to the formation of disks, jets, and outflows, and subsequently planets, has been a grand challenge. The problem contains many nonlinear couplings; e.g., the modeling of an idealized flux-frozen magnetic field leads to a complete suppression of disk formation, in contradiction to the commonly observed outcome.   Canadians have already developed various pieces of a more complete scenario. For example, \cite{Banerjee06} used idealized 3D MHD calculations to investigate the driving mechanism of low-velocity outflows and high-velocity jets during core collapse.  Also, \cite{Dapp10} and \cite{Dapp12} used thin-disk simulations with detailed chemical evolution and nonideal MHD to show that magnetic field dissipation would effectively allow disk formation. A full solution requires both 3D and nonideal MHD, and first steps have been taken by \cite{Machida19} who modeled the first two thousand years after protostar formation with self-consistent disk and jet/outflow formation.  The coming decade requires resources and collaboration to push such modeling into later phases of YSO evolution.

\subsection{How well do we understand dust within star-forming molecular clouds?}

Using continuum emission from dust to trace the mass within molecular clouds depends completely on our understanding of the optical properties of the dust itself.  
At present, we have rough estimates based on analytic models which provide uncertainties on derived object masses of perhaps half an order of magnitude.
The dust within molecular clouds may have significant differences from that within the ambient ISM, e.g., having larger sizes as a result of the colder temperatures and higher densities in clouds.  Improved understanding of the optical properties of dust will yield improved estimates of the dynamical states of structures within molecular clouds,
such as 
the ability of filaments to fragment or starless cores to collapse into new generations of young stars.  Following the low-resolution example of {\it Planck} over the past decade, better understanding of dust within molecular clouds can be attained through wide sampling of the spectral energy distributions of dust itself from near-infrared to centimeter wavelengths.  Notably, comparisons of such data with those of dust extinction made from wide-field data at visible wavelengths that will be available in the next decade will be also very important for constraining the optical properties of dust \citep[e.g.,][]{Webb17}.  Also, such wide spectral coverage will enable better constraints to be made of dust populations that are not well understood, such as emission from spinning very small grains that may account for an anomalous bump of emission at $\sim$30 GHz.  Furthermore, improved understanding of dust will allow better use of polarized continuum data to trace the magnetic fields within clouds, cores, and disks.  
  
\subsection{How much does star formation phenomena change in real time?}

Though there have been notable counterexamples (e.g., FU Ori star bursts), the previous decades of star formation research have largely regarded the processes involved in forming stars as not changing substantially on human timescales.  As a result, observed behaviour at one epoch is largely assumed to be the same behaviour at any other.  In the last decade, however, there have been increasing examples of significant short-term variability involved in star formation.  For example, the luminosities of some YSOs observed by {\it Spitzer} suggest some fraction is unusually bright in the near-infrared, likely due to enhanced short-term accretion events \citep{Dunham10}.  Moreover, variations of submillimetre intensity detected using SCUBA-2 or ALMA show that variations in the luminosities of protostars affect the heating of surrounding dust of the disk and core.  ALMA has also detected instances of chemical variations within disks as a result of disk heating and subsequent cooling \citep{Jorgensen15}.  Over the next decade, further evidence of variability in star formation will likely be captured.  First, current low-resolution surveys for periodic, episodic, and spontaneous variability will continue, providing longer temporal baselines from which transient events can be seen.  Second, high-resolution observations will be able to trace secular changes in immediate protostellar environments again over long baselines, e.g., real-time evolution of jets, outflows, and disk structures.  With better appreciation of how the star formation process varies on short timescales, we will be able to make clearer inferences on what populations of objects are telling us if observed only at a single epoch. 

\subsection{What physical processes are important (only) in clustered formation environments?}

The majority of stars appear to form within clustered environments.  Much of our deepest understanding of the processes involved in forming stars, however, is based on studies of more isolated systems, as they are the simplest ones to interpret.  Clustered environments likely do enable additional physical processes than are found in isolated star-forming systems. For example, neighbouring cores may compete for accreted material from the larger cluster potential \citep[e.g.,][]{Murray12}.  The effects of radiation from young high-mass YSOs are also stronger in the cluster-forming environment, as high-mass stars appear to form exclusively in cluster environments. Within a cluster, the first generation of massive stars to `turn on' may significantly influence the properties of additional forming systems, both at the earliest stages  \citep[e.g., increasing the Jeans mass or minimum fragmentation scale;][]{Longmore11} and at later stages \citep[e.g., substantially evaporating nearby protostellar disks;][]{Mann14}.  Further progress will require an understanding of how material in clusters is assembled, e.g., quantifying the role of filaments in supplying mass and determining the role of initial core location within a cluster in the final protostellar mass, as well as the amount and role of protostellar feedback on subsequent generations of star formation.  Additionally, larger-scale processes and special environments such as around SgrA*, the supermassive black hole at the centre of our Galaxy, merit additional study \citep[e.g.,][]{Stostad15}.  

\section{Recommendations for Investment}

Access to leading facilities has allowed Canada to be a global leader in star formation research.  In the past decade, we have benefited greatly from our associations with JCMT, {\it Herschel}, and ALMA.  In this section, we examine what associations will enable Canadians to contribute to the next chapters of this field.

\subsection{Single-dish facilities}
The research done on larger scales has been possible due to access to single-dish facilities, as these (and their focal-plane instruments) have allowed wide-field mapping of star formation environments.  For example, SCUBA-2 on JCMT and SPIRE+PACS on {\it Herschel} allowed the wide-field continuum mapping that transformed our understanding of the clumps and cores within the clouds of our Galaxy.  Also, HARP on JCMT and the KFPA on the GBT enabled the important line-emission follow-ups to the continuum maps, enabling the dynamics of the dense gas to be traced.  Going forward, we identify three important facilities:

{\it 1) The Large Synoptic Survey Telescope (LSST)}: LSST will survey the sky at visual wavelengths with high enough cadence to allow millions of transient events to be detected.  Such events toward star-forming clouds will be critical for timing the followups of such events at longer wavelengths, e.g., to determine the impact of variations of accretion onto YSOs on the surrounding disks and cores.  Moreover, LSST will eventually produce an extremely deep, multi-band map of the sky that can be used to determine the column densities of clouds at high resolution via extinction.  Such data can be critical for constraining the optical properties of dust, especially when compared with those determined at longer wavelengths \citep[e.g.,][]{Webb17}. 

{\it 2)	Ground-based submillimetre telescopes}: Canada’s success in star formation so far has been the result of access to 
the 15-m JCMT.  Going forward, access to it and other single-dish telescopes is critical.~  For example, the JCMT is pursuing a third-generation large bolometric camera, which will enable wider coverage of star-forming  clouds within the Galaxy.  Moreover, such an instrument could be used for even more extensive polarimetric mapping, enabling the role of magnetic fields within these clouds to be constrained better.  Other telescopes, however, have other promising capabilities.  For example, the 100-m diameter GBT is exploring larger-format receiver arrays in K- and W-bands that will allow very wide maps of dense gas tracers such as NH$_{3}$ and N$_{2}$H$^{+}$, respectively.  Also, the 6-m CCAT-p telescope will enable wide-field coverage of the sky from the ground at wavelengths inaccessible to the JCMT, i.e., 200-350~$\mu$m.  These activities may lead Canada to become involved in a larger, next-generation submillimetre single-dish telescope in the Atacama, e.g., CCAT or ATLAST, in the 2030s.  

{\it 3)	Trans-atmospheric far-infrared telescopes}: Building off of the successes of BLAST-pol and {\it Herschel}, Canadian access to far-infrared telescopes above Earth's atmosphere is also crucial.  For example, BLAST-TNG is a new balloon-borne mission  led by NASA with Canadian involvement to map dust polarization at resolutions 5$\times$ better than those of BLAST-pol and with mapping speeds $>$10$\times$ better. In addition, {\it SPICA} is a ESA-JAXA far-infrared 2.5-m telescope with possible Canadian involvement. Though smaller than {\it Herschel}, an actively cooled aperture will give {\it SPICA} sensitivities many orders of magnitude better.  This capability will enable much deeper censuses of YSOs and cores in molecular clouds within our Galaxy.  Moreover, the Canadian-designed FTS that is the heart of the SAFARI instrument of {\it SPICA} will be very important for constraining the optical properties of dust via wide spectral coverage of thermal emission.  Finally, the B-BOP polarimeter of {\it SPICA} will enable the kinds of wide-field maps of polarized emission needed to probe the magnetic fields associated with star formation from the scales of the cloud down to the core.  Though {\it SPICA} will not launch until the 2030s, Canadian participation in the next decade is needed to ensure leadership.  Another possibility is the NASA {\it Origins Space Telescope}, a far-infrared telescope concept, which could be conceivably larger in aperture than {\it SPICA}.  \emph{Origins}, however, remains relatively undefined at present and no timeframe for launch has been yet identified.  

\subsection{Interferometric Facilities}

Small-scale studies of star-formation in the immediate environments of cores and YSOs by Canadians have been made possible due to access to the newest, most powerful interferometric arrays, e.g., ALMA.  In the next decade, we identify three facilities where future advances in star formation research will be made.

{\it 1)	ALMA}: Though ALMA already exists, the observatory has placed significant emphasis on upgrading its capabilities over the next ten years to keep itself at the forefront of submillimetre astronomy.  For star formation research in particular, increases in the bandwidths of its various receivers and the capacity of its correlator are especially needed.  For example, increased bandwidth will allow improved sensitivity to both regular and polarized continuum observations, increasing either the speed of observing or the sizes of samples observed in a given amount of time.  Moreover, the increased bandwidth will allow more lines to be observed simultaneously with a given correlator setup, increasing the amounts of data available for studies of dynamics (and astrochemistry) on small scales.  An upgraded correlator would also allow more lines to be observed simultaneously, and also at higher spectral resolution to improve the ability of the lines to trace kinematic signatures. 

{\it 2)	ngVLA}: Canada has been a major user of the VLA over its lifetime, and plans are afoot to replace it with a new facility that will provide order of magnitude improvements in resolution and sensitivity, as well as new access at frequencies up to 115 GHz.  The ngVLA will have a major impact on star formation in several ways.  For continuum data, its high sensitivity will enable probes of emission from star-forming dust at much lower optical depths than possible even at submillimetre wavelengths, e.g., disks or high-mass protostellar cores.   Also, its polarization capabilities will enable magnetic fields in these environments to be better discerned.  For line data, the ngVLA will allow much wider-field mapping of key tracers of dense gas in star-forming environments such as NH$_{3}$ and N$_{2}$H$^{+}$ emission than currently possible.  In addition, there will be better capabilities to probe key tracers of magnetic fields in clouds via the Zeeman effect on lines such as those of OH and CCS.  

{\it 3)	SKA}: Canada is a partner in the SKA project and is making major technical and scientific contributions.
The most important contributions to Galactic star formation are likely to come from observations at the high-frequency side of the SKA1 range. These include spatially resolving the growth of dust grains in protostellar disks as planetary system formation begins. The formation of heavy organic molecules is expected to be very efficient on the surfaces of dust grains, which then can be liberated into the gas phase by cosmic rays and irradiation by the central protostar or by external radiation fields in high-mass YSO environments. The rotational transitions of these heavy molecules occur at the relatively low frequencies observable by SKA1. Observations 
with SKA1 will also image regions near protostars where jets are launched and collimated in free-free or HI 21-cm emission. 

 




~

\begin{lrptextbox}[How does the proposed initiative result in fundamental or transformational advances in our understanding of the Universe?]\vspace{-0.3cm}
Star formation impacts a wide range of fields from planets to galaxies.
\end{lrptextbox}

\begin{lrptextbox}[What are the main scientific risks and how will they be mitigated?]
\vspace{-0.3cm}
Involvement in no new facilities would limit our ability to make progress in the field.
\end{lrptextbox}

\begin{lrptextbox}[Is there the expectation of and capacity for Canadian scientific, technical or strategic leadership?] 
\vspace{-0.3cm}
Yes, Canada continues to play a leadership role in many of the science and technical advances in the field.
\end{lrptextbox}

\begin{lrptextbox}[Is there support from, involvement from, and coordination within the relevant Canadian community and more broadly?] 
\vspace{-0.3cm}
This WP is the collaborative effort of the Canadian community.  All large new initiatives are multi-national.
\end{lrptextbox}

\begin{lrptextbox}[Will this program position Canadian astronomy for future opportunities and returns in 2020-2030 or beyond 2030?] 
\vspace{-0.3cm}
Involvement in at least some of the facilities described will keep Canada competitive in future developments.
\end{lrptextbox}

\begin{lrptextbox}[In what ways is the cost-benefit ratio, including existing investments and future operating costs, favourable?] 
\vspace{-0.3cm}
The current and proposed facilities benefit a wide range of science cases, not just star formation.
\end{lrptextbox}

\begin{lrptextbox}[What are the main programmatic risks
and how will they be mitigated?] \vspace{-0.3cm}
Investment in new facilities all require multi-national agreements.  Canada continues to investigate contributions across a variety of facilities to reduce the risk of none succeeding.

\end{lrptextbox}

\begin{lrptextbox}[Does the proposed initiative offer specific tangible benefits to Canadians, including but not limited to interdisciplinary research, industry opportunities, HQP training,
EDI,
outreach or education?] 
\vspace{-0.3cm}
Star formation has a notably diverse researcher population, while outreach is aided by our spectacular images and accessible topics.  Industrial partnerships are possible for many of the proposed facilities.
\end{lrptextbox}

\setlength{\bibsep}{0pt plus 0.3ex} 
\renewcommand{\baselinestretch}{0.9} 
\twocolumn{
\footnotesize{
\bibliography{example}
}}
\end{document}